\documentclass[
 reprint,
 amsmath,amssymb,
 aps,
pra,
]{revtex4-2}

\usepackage{graphicx}
\usepackage{dcolumn}
\usepackage{bm}
\usepackage{xcolor}
\usepackage{quantikz}
\usepackage{circuitikz}
\usepackage{natbib}  
\usepackage{float}
\usepackage{epstopdf} 
\usepackage{rotating}
\usepackage{longtable}
\usepackage{amsmath}
\usepackage{hyperref}

\begin{document}
\newcommand{\mE}{\mathcal{E}}
\newcommand{\WH}[1]{\textcolor{blue}{\textbf{WenHan:} #1}}
\newcommand{\V}[1]{\textcolor{red}{\textbf{Valerio:} #1}}
\newcommand{\C}[1]{\textcolor{orange}{ #1}}

\preprint{APS/123-QED}

\title{Petz recovery maps of single-qubit decoherence channels in an ion trap quantum processor}

\author{Wen-Han Png$^1$}
\author{Valerio Scarani$^{1,2}$}
\affiliation{$^1$Centre for Quantum Technologies, National University of Singapore, 3 Science Drive 2, Singapore 117543}
\affiliation{$^2$Department of Physics, National University of Singapore, 2 Science Drive 3, Singapore 117542}

\date{\today}

\begin{abstract}
The Petz recovery map provides a near-optimal reversal of quantum noise, yet proposals for its implementation are only recent. We propose a physical realization of the exact state-specific Petz map in an ion trap for qubit decoherence channels. Our circuit constructions require at most $1 (2)$ ancilla qubits and $3 (20)$ CNOT gates for channels with Kraus rank $2 (>2)$. We analyze typical ion trap errors and construct corresponding Petz maps, simulating their performance under realistic noise modeled by residual spin-motion coupling. Quantum circuits are provided for depolarizing, dephasing, and amplitude damping channels. Focusing on single-shot recovery, suited for present-day devices, we also quantify the precision of prior knowledge required to achieve a recovery error below 0.01 across varying decoherence levels and state purities.

\end{abstract}

\maketitle


\section{\label{sec:Intro}Introduction}

Quantum systems are never truly isolated. Processes and protocols are done in the presence of environments that introduce noise. As such, recovery protocols that can undo the unintended dynamics are of great utility. Petz's theory of statistical sufficiency \cite{Petz1986sufficient} defines a transpose channel $\mathcal{R}_{(\mE,\gamma)}$ (also known as Petz recovery map, defined for a quantum channel $\mE$ and reference state $\gamma$) that recover a quantum state affected by noise \cite{junge2018universal, wilde2013quantum}. The Petz map can be understood as a quantum generalization of Bayes' theorem \cite{leifer2013towards,parzygnat2023axioms,parzygnat2023time,cenxin2023quantum}. As a brief intorduction, Bayes' theorem updates a prior distribution $ \gamma(\theta) $ over parameters $\theta$ based on observed data $x$, with the posterior distribution given by:
\begin{equation}
    \hat{\varphi}_{\gamma}(\theta|x) = \frac{\varphi(x|\theta)\gamma(\theta)}{\varphi[\gamma](x)}
\end{equation}
Here, $\varphi(x|\theta)$ is the likelihood, and $\varphi[\gamma](x)=\sum_{\theta}\varphi(x|\theta)\gamma(\theta)$ is the normalization factor. From an information-theoretic perspective, Bayes' theorem acts as a reversal of the likelihood function $\varphi(x|\theta)$, reconstructing the distribution over $\theta$ based on given data. As we shall describe in detail later, the Petz recovery map \( \mathcal{R}_{\mathcal{E},\gamma} \) serves as the quantum analog, where the likelihood corresponds to a quantum channel $\mE$ and the prior is represented by a density matrix $\gamma$. Here, $\gamma$ is also often referred to as a reference state. It is not a physical quantum state, rather it represents the prior knowledge we have for the input state before applied with $\mE$. It is also the state that we choose to recover perfectly by applying the Petz map on the output of the channel: indeed, $\mathcal{R}_{\mathcal{E},\gamma}[\mE[\gamma]]=\gamma$.

The Petz recovery map is often used in quantum information as a theoretical tool, primarily for checking to what extent recovery from noise is possible. The performance, dependence on reference state, and dimensionality of the Petz recovery map has been investigated for typical quantum channels such as dephasing and amplitude damping \cite{LAUTENBACHER2024129583}. Notably, it is known that the map is near-optimal in reversing the effect of noise \cite{barnum2002reversing}. It has also been shown as a universal recovery operation for approximate quantum error correction, which is once again near-optimal in terms of worst-case fidelity \cite{ng2010simple,tyson2010two,mandayam2012towards}. It has also been applied in the decoding process to obtain a coherent information rate in quantum communication channels
\cite{beigi2016decoding}, with further applications in quantum channel capacities \cite{hausladen1996classical,holevo1998capacity,schumacher1997sending,beigi2014quantum}, state discrimination \cite{barnum2002reversing,belavkin1975optimal,belavkin1975optimal2,holevo1978asymptotically}, and entanglement wedge reconstruction \cite{chen2020entanglement}.

Recent studies have proposed circuit implementations of the Petz map. In the first \cite{gilyen2022quantum}, the state-specific Petz map \cite{barnum2002reversing} was described using block encoding and quantum singular value transformation. An exact implementation of the code-specific Petz map \cite{ng2010simple} was then proposed in \cite{Biswas2024}. There, the authors exploited the fact that the Petz map is a CPTP map, and as such can be isometrically extended to a unitary circuit by Stinespring dilation. Besides being exact, the implementation of the isometric extension is also better in its scaling with the number of qubits to be corrected. 

In this work, we take a further step towards practicality and describe the implementation of Petz recovery maps for a trapped-ion qubit and its most frequent decoherence channels. The map will be implemented 
using geometric phase gates as the interaction with the ancilla. We analyze the error thresholds required for the geometric phase gate to ensure reliable operation.


This paper is structured as follows, In Sec.~\ref{sec:Petz} we introduce the Petz recovery map and its circuit implementation for arbitrary-rank decoherence channel affecting a single qubit; then we examine the requirements for the prior distribution of the reference state to ensure that the recovery error is bounded below 0.01. In Sec.~\ref{sec:ionAll}, we describe the trapped ion system and common error sources that give rise to dephasing, amplitude damping, and depolarization. For each of these errors, we determine the error thresholds of the geometric phase gate needed to realize the corresponding Petz recovery maps.

\section{\label{sec:Petz}Petz recovery map}

In this section, we briefly introduce the Petz recovery map. To begin with, we consider a forward channel $\mE$ acting on a qubit input, where $\mE$ is a CPTP map of rank $M$ given by $\mE(\bullet)=\sum_{m=1}^{M}E_{m}\bullet E_{m}^{\dag}$. For a qubit channel, $\bullet \in \mathbb{C}^2$ and $M = {1,...,4}$. One can regard $\mE$ as a channel that introduces undesired dynamics, commonly regarded as noise or decoherence. The question of whether the noise can be reversed was first addressed in \cite{Petz1986sufficient}. As a consequence of statistical sufficiency in Petz's theory, a decohered state $\mE(\rho)$ can be fully recovered, e.g. $\hat{\mathcal{R}}_{(\mE,\gamma)}\circ \mE(\rho) = \rho$  if and only if
\begin{equation}
    D(\mE(\rho)|\mE(\gamma))=D(\rho|\gamma) \label{Eq:PerfectRecovery}
\end{equation}
where  $D(\rho|\gamma)= \text{Tr}[\rho(\log\rho-\log\gamma)]$ is the Umegaki quantum relative entropy of the two states. Given $\mE$ and a reference state $\gamma$, the corresponding transpose channel of $\mE$ can be constructed as follows
\begin{equation}
    \hat{\mathcal{R}}_{(\mE,\gamma)}(\bullet) =\sqrt{\gamma}\mE^\dag\left(\frac{1}{\sqrt{\mE(\gamma)}}\bullet\frac{1}{\sqrt{\mE(\gamma)}}\right)\sqrt{\gamma}\label{Eq:Petz}
\end{equation}
In Kraus representation, the Petz map read
\begin{equation}
    \hat{\mathcal{R}}_{(\mE,\gamma)}(\bullet) = \sum_{m=1}^{M} K_{m}^{(\mE,\gamma)} \bullet K_m^{\dag(\mE,\gamma)} \label{Eq:PetzKraus}
\end{equation}
where $K_{m}^{(\mE,\gamma)} = \sqrt{\gamma} E_m^\dag \frac{1}{\sqrt{\mE(\gamma)}}$. The recovery fidelity of Petz map is given by $\mathcal{F}(\rho,\hat{\mathcal{R}}_{(\mE,\gamma)}\circ \mE(\rho))$, where the fidelity is defined as $\mathcal{F} (\rho_0,\rho_1) = \text{Tr}\Big[\sqrt{\sqrt{\rho_0}\rho_1\sqrt{\rho_0}}\Big]^2$.

\subsection{\label{sec:Iso} Isometric extension for the Petz Map of a single-qubit decoherence channel}

A Petz recovery map is a CPTP map in general $ \hat{\mathcal{R}}_{(\mE,\gamma)}  : \mathcal{S}( \mathcal {H}_A) \rightarrow \mathcal{S}( \mathcal {H}_C)$. We define the ancillary system as $\rho_{B} \in \mathcal{S}(\mathcal{H}_B)$ where $\mathcal{H}_B$ is the Hilbert space of the ancillary system. The isometric extension of the $\hat{\mathcal{R}}_{(\mE,\gamma)}$ can be constructed by defining a unitary channel $U_{(\mE,\gamma)} \bullet U_{(\mE,\gamma)}^\dag :  \mathcal{S}( \mathcal{H}_B \otimes \mathcal {H}_A ) \rightarrow \mathcal{S}( \mathcal {K}), \mathcal {K} = \mathbb{C}^{2M}$, followed by tracing the ancillary system
\begin{equation}
    \hat{\mathcal{R}}_{(\mE,\gamma)}(\bullet)=\text{Tr}_{B}\left[U_{(\mE,\gamma)}(\rho_{B}\otimes \bullet)U_{ (\mE,\gamma)}^{\dag}\right].
\end{equation}

Using the trace-preserving (TP) condition $\sum_{m=1}^{M}K_{m}^{ \dag(\mE,\gamma)} K_{m}^{(\mE,\gamma)}  =\mathcal{I}_{A}$, the sum of each diagonal entry gives 1 each, while the sum of each off-diagonal term gives zero. Let $\rho_{B} = |0\rangle\langle0|$,
the TP condition allows us to construct $U_{(\mE,\gamma)}$  using Stinespring's dilation
\begin{equation}
    U_{(\mE,\gamma)}
=\left[\begin{array}{cccc}
 K_{1}^{(\mE,\gamma)} & . & . & .\\
 K_{2}^{(\mE,\gamma)} & . & . & .\\
\vdots & . & . & .\\
 K_{M}^{(\mE,\gamma)} & . & . & .
\end{array}\right] \label{Eq:GS}
\end{equation}
where the first two columns are mutually orthogonal. Here, $U_{(\mE,\gamma)} \in U(2M)/U(1)$ up to a global phase. The rest of the column of $U_{(\mE,\gamma)}$ can be constructed using Gram-Schmidt orthogonalisation. 

Next, we discuss the resources required for  circuit implementation of $U_{(\mE,\gamma)}$ in terms of number of ancillary qubit and number of CNOT gates. For the rest of the paper,  we only consider the Petz recovery map for single-qubit decoherence. In this case, the dimension of the ancillary system for a Petz map is given by $M$. Straightforwardly, for $M=2$ and $M>2$, the Petz map requires one ancilla qubit and two ancilla qubits respectively. 
For $M=2$, $U_{(\mE,\gamma)}$ belongs to $SU(4)$. Therefore, it can always be decomposed into at most three CNOT gates and single qubit gates \cite{wang2021single,harty2014high}. For  $M=3,4$ , the maximum CNOT gates required is $20$ \cite{shende2004minimal}.

\subsection{\label{sec:Prior} Choice of reference state and recovery fidelity }

\begin{figure}
    \centering
    \includegraphics[width=0.9\linewidth]{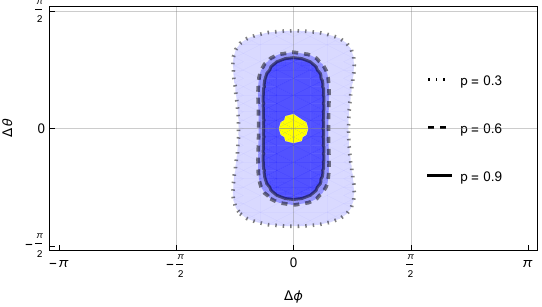}
    \includegraphics[width=0.9\linewidth]{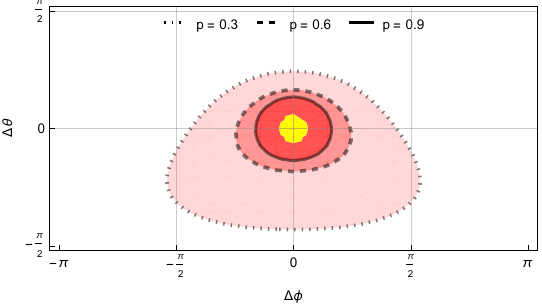}
    \includegraphics[width=0.9\linewidth]{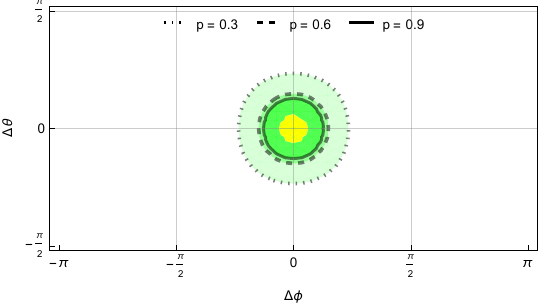}
    \caption{Parametric plot of $ \delta \mathcal{F} =0.01$ in the 2 dimensional parameter space of $(\Delta \phi,\Delta \theta)$ with $(\theta_0 = \pi/2 , \phi_0 = \pi/4 )$ fixed $R_0 = 0.5$, and  $\Delta R =0$. The figure from top to down are plots for dephasing (blue), amplitude damping (red) and depolarising (green) respectively. The shading indicates the region where $\mathcal{F} \leq 0.01$. The region shrinks with increasing $p$. }
    \label{fig:FidVsP0}
\end{figure}

\begin{figure}
    \centering
    \includegraphics[width=0.9\linewidth]{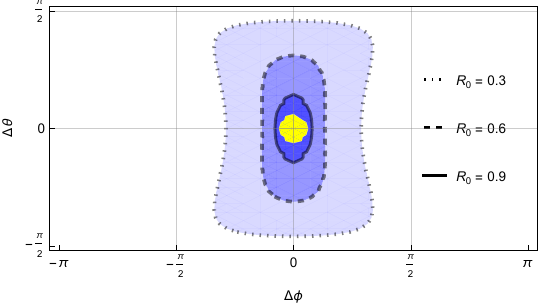}
    \includegraphics[width=0.9\linewidth]{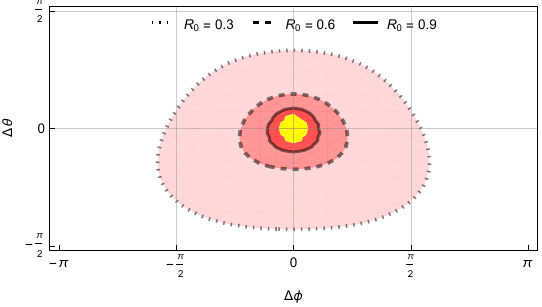}
    \includegraphics[width=0.9\linewidth]{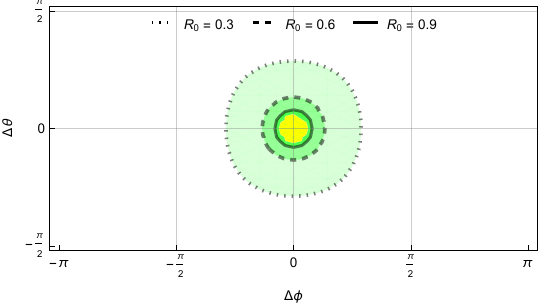}
    \caption{Parametric plot of $ \delta \mathcal{F} =0.01$ in the 2 dimensional parameter space of $(\Delta \phi,\Delta \theta)$ with $(\theta_0 = \pi/2 , \phi_0 = \pi/4 )$ fixed $p = 0.5$, and  $\Delta R =0$. The figure from top to down are plots for dephasing (blue), amplitude damping (red) and depolarising (green) respectively. The shading indicates the region where $\mathcal{F} \leq 0.01$. The region shrinks with increasing $R$. }
    \label{fig:FidVsR0}
\end{figure}

The perfect recovery condition in Eq.~\eqref{Eq:PerfectRecovery} is generally unrealistic, as exact information about a state before decoherence is rarely available. Instead, one typically has only prior information about the reference state, leading to an approximate recovery with a finite error. This prior information can be obtained from previous experimental data. For a single qubit, we specifically examine the minimal deviation from the condition in Eq.~\eqref{Eq:PerfectRecovery} that ensures the recovery error is bounded below 0.01, defining a sufficiently good recovery in a fault-tolerant regime \cite{shor199637th,benhelm2008towards}. To formulate the problem, we consider the following setting where $\rho$ is the state before the decoherence $\mE$, and $\gamma$ is the reference state. Using the parameters of Bloch sphere, we define $\rho = \rho (R_0, \theta_0, \phi_0)$. The recovery error for a given input state $\rho$ is given by
\begin{equation}
    \delta \mathcal{F} (\Delta R, \Delta \theta, \Delta \phi) = 1-\mathcal{F}(\rho,\hat{\mathcal{R}}_{(\mE,\gamma)}\circ \mE(\rho))
\end{equation}
where $\rho = \rho(R_0, \theta_0, \phi_0)$ and $\gamma = \rho(R_0 + \Delta R, \theta_0 + \Delta \theta,\phi_0 + \Delta \phi )$. Here, the tuples $(\Delta R, \Delta \theta, \Delta \phi)$ represent our ignorance on the state $\rho$. After fully characterized the error $\mE$, we can define a Petz recovery based on $\gamma$. For non-zero tuples $(\Delta R, \Delta \theta, \Delta \phi)$, we can only get partial recovery, given by $\delta \mathcal{F}>0$.  $\delta \mathcal{F} $ in general spans a 4-dimensional parameter space of $(p, \Delta R, \Delta \theta, \Delta \phi)$ where $p\in [0,1]$ defines the degree of decoherence. We present our analysis in the 2D parameter space of $(\Delta\theta,\Delta\phi)$ which is relative to $(\theta_0,\phi_0)$ (see Figures \ref{fig:FidVsP0} and \ref{fig:FidVsR0}). The region from the origin to the contour line with $\delta \mathcal{F} = 0.01$ defines the area of the prior distribution of $\gamma$ where $ \delta \mathcal{F} \leq 0.01$. The area defines how stringent the prior information we should have for a reference state for a recovery up to 0.01.

We then examine how the region bounded changes in the two cases (i) fixed $R_0 = 0.5$, $p = \{ 0.3, 0.6, 0.9 \}$ and (ii) fixed $p_0 = 0.5$,  $R_0 = \{ 0.3, 0.6, 0.9\}$.  We observed that the boundary for $\delta \mathcal{F} = 0.01$ shrinks as $R_0$ and $p$ increase. This suggests that recovering a state with high purity after severe decoherence generally requires precise knowledge of the reference state. These results provide guidance for preparing the prior information, which aids in the first step of quantum Bayesian inference.  In example of $(\theta_0 = \pi/2 ,\phi_0= \pi/4 )$ , we numerically compute the boundary for dephasing channel, amplitude damping channel, and depolarising channel [see Fig.~\ref{fig:FidVsP0} for case (i) and Fig.~\ref{fig:FidVsR0} for case (ii)]. 
In the extreme case where $R_0 \rightarrow 1$ and $p \rightarrow 1$, the Petz recovery map for an erasure channel reconstructs the exact pure reference state, such that $\hat{\mathcal{R}}_{(\mathcal{E}, \gamma)} \circ \mathcal{E}(\rho) = \gamma$. The recovery error expression becomes $\delta \mathcal{F} (\Delta R, \Delta \theta, \Delta \phi) = 1 - \mathcal{F}(\rho, \gamma)$. The distribution of $(\Delta \phi, \Delta \theta)$ under these extreme conditions is represented by the innermost yellow region in Figures \ref{fig:FidVsP0} and \ref{fig:FidVsR0}. For more examples, such as $(\theta_0 = \pi/4, \phi_0 = \pi/4)$, refer to Appendix \ref{sec:PriorBoundary}.

\section{\label{sec:ionAll} Implementation in trapped ions information processor}

One of the primary decoherence channels in trapped ions is dephasing, typically caused by control laser intensity fluctuations, magnetic field fluctuations, laser frequency noise, and ambient electric field noise \cite{wang2021single,harty2014high}. Another common decoherence mechanism is amplitude damping, arising from energy relaxation. 
For an optical qubit, spontaneous emission of the metastable qubit level to the ground qubit level leads to amplitude damping \cite{staanum2004lifetime,nigg2013experimental,schindler2013quantum,clark2021high}; remaining spontaneous emission leads to leakage, which is beyond the scope of this paper. While hyperfine and Zeeman qubits are relatively stable, creating and manipulating entangled state make the qubits prone to off-resonant scattering. When the scattering rates to the ground and excited states are different, amplitude damping arises. Thus, even when the dephasing error is suppressed, the amplitude damping time $T_1$ will limit the coherence time of an ion qubit.   Qubit decoherence processes characterized by Kraus rank greater than two generally correspond to a non-trivial mixture of distinct error mechanisms. In particular,  randomized benchmarking experiments with trapped ions often approximate the overall gate errors of the circuit as a depolarizing channel \cite{knill2008randomized,gaebler2016high,ballance2016high}. In the following section, we show the implementation of Petz recovery map in trapped ions system for the dephasing, amplitude damping and depolarising channels. 

Before we do so, we give a quick review of the trapped ions system \cite{bruzewicz2019trapped}. An ion trap can be described through the joint system of $N$ two-level system (qubit) and the $3N$ mode quantum harmonic oscillator. A qubit is encoded in two electronic states of an ion. The quantum harmonic oscillator is the quantized version of effective Coulomb interaction between the ions, also known as the motional state. The effective Coulomb interaction takes into account the harmonic electric confinement potential and the mutual Coulomb interaction of the ions, which can be transformed into normal mode coordinates that describe collective motional state. In most of the quantum applications, it is suffices to couple the ions to the motional state along one mode.

\subsection{\label{sec:ionIntro} Ion-motion coupling and geometric phase gate}

We consider the Hamiltonian of on the spin-dependent motional coupling given as \cite{lee2005phase, leibfried2003experimental} 

\begin{equation}
    \tilde{H}_{int}(t)=\left[A^{*}(t)\hat{a}-A(t)\hat{a}^{\dagger}\right]\hat{J}_{z} \label{Eq:TrappedIonHamiltonian}
\end{equation}
where $\hat{J}_{z}=\hat{\sigma}_{z}^{(A)}+\hat{\sigma}_{z}^{(B)}$, $A(t)=ge^{i\delta t}$. Here, $g$ is the spin-motion coupling strength, $\delta=\omega-\omega_{0}$ is the detuning between the laser frequency and the motional frequency.
We have assumed that the laser imprints the same $g$ and $\delta$ on both ions. The time-ordered evolution of Eq.~\eqref{Eq:TrappedIonHamiltonian} gives 
\begin{equation}
\mathcal{T}\exp\left(-i\int_{0}^{t}\tilde{H}(t')dt'\right) 
 =D(\alpha(t)\hat{J}_{z}) U_{ZZ}(\Phi(t))  \label{Eq:evo}
\end{equation}
where $D(\beta) = e^{\beta a^\dagger - \beta^* a}$ represents a residual spin-dependent motional displacement, and $U_{ZZ}(\theta) = \exp\left(-i \theta \hat{\sigma}_{z}^{(A)}\hat{\sigma}_{z}^{(B)}\right)$ is the geometric phase gate. The amount of spin-dependent motional excitation is given by the trajectory in the phase space $\alpha(t)=-i\frac{g}{\delta}(e^{i\delta t}-1)$, while the phase accumulated by spin-spin coupling is given by the area of the closed trajectory in the phase space $\Phi(t) =\int_{0}^{t}dt'\Im\left[A(t')\alpha^{*}(t')\right]=(\frac{g}{\delta})^2\left(\delta t-\sin\delta t \right)$. When $t = T = 2 L \pi/\delta, L\in \mathbb{Z}$, one can see $\alpha(T)=0$ and we obtain the ideal geometric phase gate with $\Phi(T)=\frac{g^2}{\delta}T$. This serves as a primitive gate for a two-qubit gate in trapped ions quantum computers.

The noise in the geometric phase gate can be characterized as the perturbation of the spin-motion coupling strength. Since $\alpha(t)\sim O(g/\delta)$ and $\Phi(t)\sim O\left((g/\delta)^2 )\right)$, we take the two-qubit gate error as $\Delta =( \frac{\epsilon_g}{\delta})^{2}$ where $\epsilon_g$ is the offset of the laser coupling strength;
whence the channel of the noisy geometric phase gate is 
\begin{equation}
    \mE_{GPG}^{(\Delta)}(\bullet)=\mE_{ph}^{(\Delta)}\circ\mathcal{U}_{sys}^{(\Delta)}\circ \mathcal{U}_{ZZ}(\bullet)
\end{equation}
where $\bullet\in\mathcal{H}_{A}\otimes\mathcal{H}_{B}$ denotes the qubit-ancila system. Here,  $\mathcal{U}_{ZZ}(\bullet)  = U_{ZZ}^{\dag}(\Phi(T))\bullet U_{ZZ}(\Phi(T))$ is the ideal geometric phase gate, $\mathcal{U}_{sys}^{(\Delta)}(\bullet)  = U_{ZZ}^{\dagger}(\Delta)\bullet U_{ZZ}(\Delta)$ is the erroneous spin-spin excitation and $\mE_{ph}^{(\Delta)}$ is the phase-flip channel due to residual spin-motion coupling.    $\mE_{ph}^{(\Delta)}(\bullet)=E_{0}\bullet E_{0}^{\dag}+E_{1}\bullet E_{1}^{\dag}$, where $E_{1}=e^{-|\sqrt{\Delta}|^{2}/2}\sqrt{\cosh(\Delta)}I$ and $
E_{2}=e^{-|\sqrt{\Delta}|^{2}/2}\sqrt{\sinh(\Delta)}CZ
$. Here $CZ = \text{diag} (1,1,1,-1)$ denotes the controlled-$Z$ operation (see Appendix \ref{sec:ResidualDephase} for the derivation). On can quickly check that ideal geometric phase gate is recovered for $\Delta =0$.

\subsection{\label{sec:imp}Implementation of Petz map}

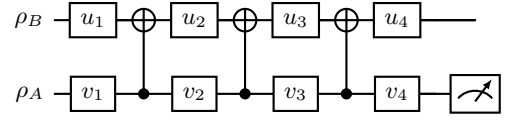
\begin{figure}
\centering
\begin{quantikz}[column sep=0.2cm]
    \lstick{$\rho_B$}  &\gate{u_1} & \targ &\qw &\gate{u_2} &\targ  &\qw &\gate{u_3}  &\targ &\qw  &\gate{u_4} &\qw &\qw  \\
    \lstick{$\rho_A$}  &\gate{v_1} & \ctrl{-1} &\gate{v_2}  &\ctrl{-1} &\gate{v_3} &\ctrl{-1} &\gate{v_4} &\qw &\meter{} 
\end{quantikz}
\caption{Gate decomposition of any $SU(4)$ unitary for decoherence channel with Kraus rank-2,where $u_l\in SU(2)$ and $v_l\in SU(2)$  \cite{Vidal2004} }
\label{Fig:Circuit} 
\end{figure}

The implementation of a Petz map can be divided into two steps:  construction of the Petz unitary $U_{(\mE,\gamma)}$, and circuit decomposition.
Before the implementation, state tomography on the state before decoherence is crucial to get the information on the reference state $\gamma$ . Besides, error characterization \cite{bruzewicz2019trapped,haffner2008quantum} on the decohered state is also required to obtain the degree of decoherence $p$. Both $\gamma$  and $p$ for given channel $\mE$ together defines a Petz unitary.
We note that the circuit decomposition can be performed numerically using a quantum compiler package \cite{iten2016quantum}. Thus, we focus on the construction of the Petz unitary $U_{(\mE,\gamma)}$, specifically for a decoherence channel of Kraus rank 2.

With full knowledge of $\mathcal{E}$ and $\gamma$, one can derive the Kraus operators of the Petz map using Eq.~\eqref{Eq:PetzKraus}. The Petz map of dephasing channel has Kraus operators $K_0^{(\mE,\gamma)} = \sqrt{\gamma}\frac{1}{\sqrt{\mE(\gamma)}}\sqrt{1-p/2}$  and $K_1^{(\mE,\gamma)} =\sqrt{\gamma}\sigma_{z}\frac{1}{\sqrt{\mE(\gamma)}}\sqrt{p/2}$. While for amplitude damping channel, the Kraus operators are $K_0^{(\mE,\gamma)} = \sqrt{\gamma}(|0\rangle\langle 0| + \sqrt{1-p} |1\rangle\langle 1 |)\frac{1}{\sqrt{\mE(\gamma)}}\sqrt{p}$ and $K_1^{(\mE,\gamma)} =\sqrt{\gamma}  (|1\rangle\langle 0| \sqrt{p}) \frac{1}{\sqrt{\mE(\gamma)}}$. Then, we apply singular value decomposition (SVD) to the Kraus operator $K_{m}^{(\mE,\gamma)}=U_{m}D_{m}V_{m}^{\dagger}$, where $U_{m},V_{m}$ are unitary matrices and $D_m = \text{diag}(a_m,b_m)$ is a diagonal matrix.  $a_{m}$ and $b_m$ are the eigenvalues of $K_{m}^{(\mE,\gamma)}$. Here, the matrix entries of $U_{m},D_{m},V_{m}^{\dagger}$  depend on $\mE$ and $\gamma$.  Using the trace preserving property of the Petz map and unitary transformation to the basis of $D_{1}$, we have

\begin{equation}
|\tilde{D}_0|^{2}+|D_{1}|^{2}  =\mathcal{I}_{A}
\end{equation}
where $|\tilde{D}_0|^{2} = V_{1}^{\dag}V_{0}|D_{0}|^{2}V_{0}^{\dag}V_{1}$ is also a diagonal matrix with transformed diagonal values ${\tilde{a},\tilde{b}}$.  The trace-preserving relation implies $\tilde{a_{0}}+a_{1}=1$ and $\tilde{b_{0}}+b_{1}=1$. The Kraus operators after the unitary transformation to the basis of $D_{1}$ read


\begin{equation}
\begin{aligned}
    K_{0}^{(\mE,\gamma)} & = \tilde{U_{0}}\left(\tilde{D_{0}}\right)V_{1}^{\dag} \\
    K_{1}^{(\mE,\gamma)} & = U_{1}D_{1}V_{1}^{\dag}
\end{aligned}
\end{equation}
where $\tilde{D_{0}}=V_{1}^{\dag}V_{0}D_{0}V_{0}^{\dag}V_{1}$ and
$\tilde{U_{0}}=U_{0}V_{0}^{\dag}V_{1}$. Using Eq. \eqref{Eq:GS} and SVD, the Petz unitary read

\begin{equation}
    U_{(\mE,\gamma)} =\left(\begin{array}{cc}
            \tilde{U_{0}} & 0\\
            0 & U_{1}
            \end{array}\right) \mathbf{D}  \left(\begin{array}{cc}
            V_{1}^{\dag} & 0\\
            0 & G
            \end{array}\right)\\ 
\end{equation}
where $\mathbf{D} = \left(\begin{array}{cc}
\tilde{D_0} & D_1 \cdot(-i\sigma_y)\\
D_1 & \tilde{D_0} \cdot(i\sigma_y)\end{array}\right)$ is the one possible construction of orthogonal matrix and $G$  can take any arbitrary 2 by 2 unitary matrix. Since for both the dephasing and the amplitude damping channels, $U_{(\mE,\gamma)}\in SU(4)$, the circuit decomposition follows the format in Fig. \ref{Fig:Circuit}.  Here the single qubit unitaries  $u_l$ and $v_l$, $l=1,...,4$ varies with $\mE$ and $\gamma$.  We give the circuit decomposition for the dephasing channel and the amplitude damping channel (see Fig.~\ref{Fig:CircuitDephase} and Fig.~\ref{Fig:CircuitAmpDamp}) using $p=0.5$ and reference state $\gamma(R=0.5,\theta = \pi/2, \phi = \pi/4)$. 

An explicit analytical form of $U_{(\mathcal{E},\gamma)}$ for $M>2$ is generally difficult to obtain; however, it can be constructed by Gram–Schmidt orthogonalisation. The Kraus operators for the corresponding Petz recovery map can be obtained using Eq.~\eqref{Eq:PetzKraus} where we bring in the Kraus operators of depolarising channel given by $E_{0} = \sqrt{1-p},\mathcal{I}$ and $E_{i} = \sqrt{p/3} \sigma_{i}$, $i \in \{x, y, z \}$.  We provide an example of circuit realization of the depolarizing channel for $p=0.5$ and $\gamma(R=0.5,\theta = \pi/2, \phi = \pi/4)$ in Appendix \ref{sec:CircuitDepo} (see Fig. \ref{Fig:CircuitDepolarising}), where the circuit only requires two ancillary qubits and 20 CNOT gates.

\subsection{\label{sec:ionErrorThreshold}Error threshold of geometric phase gate}

\begin{figure}
    \centering
    \includegraphics[width=0.9\linewidth]{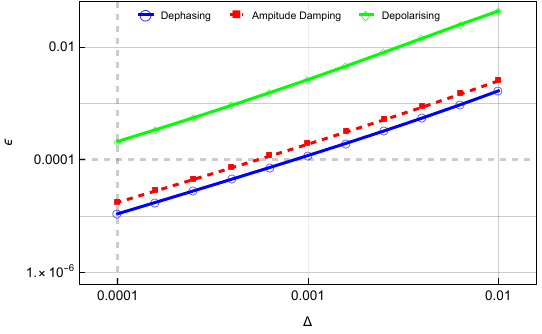}
    \caption{Plot of the average recovery error of noisy Petz recovery map against the error of geometric phase gate for  $p=0.5$. The average is taken over uniform sample of over $10^6$ points from a Bloch sphere.} 
    \label{fig:FidThresAv}
\end{figure}

\begin{figure}
    \centering
    \includegraphics[width=0.9\linewidth]{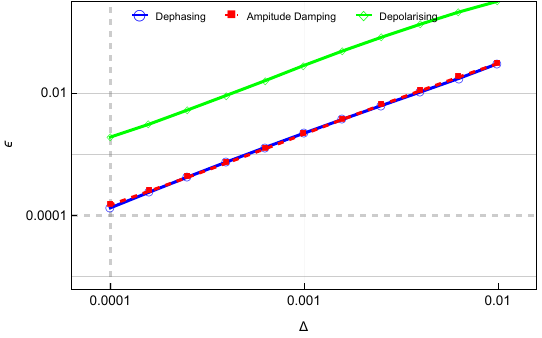}
    \caption{Plot of maximum observed recovery error of noisy Petz recovery map  against the error of geometric phase gate for  $p=0.5$. The maximum value is obtained over uniform sample of $10^6$ points from a Bloch sphere. }
    \label{fig:FidThresMax}
\end{figure}
In this section, we give the error threshold of the geometric phase gate for Petz recovery in condition Eq. \eqref{Eq:PerfectRecovery}  As discussed in Sec. \ref{sec:imp}, one can decompose $U_{(\mE,\gamma)}$ into single-qubit rotations and CNOT gates. To make it relevant to trapped ions qubits, we decompose the unitary $U_{(\mE,\gamma)}$ into single-qubit rotations and geometric phase gate gate.  Then, we define a noisy Petz map channel $\mE_{(\mE,\gamma)}^{(\Delta)}$, which is constructed by replacing $\mathcal{U}_{ZZ}$ with the noisy gate channel $\mE_{GPG}^{(\Delta)}$ 
\begin{equation}
    \hat{\mathcal{R}}_{(\mE,\gamma)}^{(\Delta)} = \text{Tr}_{B}\left[ \mE_{(\mE,\gamma)}^{(\Delta)} (\rho_{B}\otimes \bullet)\right] \label{Eq:NoisyPetz}
\end{equation}
where the noisy quantum circuit is defined as follows $\mE_{(\mE,\gamma)}^{(\Delta)} = \mE_{RM}^{(\Delta)} \circ \mathcal{U}_{(\mE,\gamma)}^{(\Delta)}(\bullet)$. Here, $\mathcal{U}_{(\mE,\gamma)}^{(\Delta)}$ gives a systematic shift to the rotation angle of both the single qubit gates and the geometric phase gate gates. $\mE_{RM}^{(\Delta)} $ gives decoherence due to residual spin-motion entanglement. For a generic quantum circuit, the effective decoherence of the overall spin-motion residual entanglement reads
\begin{equation}
    \mE_{RM}^{(\Delta)} (\bullet)= \text{Tr}_{B}\left[\Pi_{l=1}^{n}D(\sqrt{\Delta_{l}}\hat{G}_{l})(\bullet\otimes|0\rangle\langle0|)D(-\sqrt{\Delta_{l}}\hat{G}_{l})\right]
\end{equation}
where $n$ denotes number of geometric phase gate gates and $\hat{G}=\sum_{i,j}\phi_{i,j}\hat{S}_{i}\otimes\hat{S}_{j}$
is a Hermitian operator that serves as a generator for two-qubit operations. Here $\hat{S}_{i}=\{I,\sigma_{x},\sigma_{y,}\sigma_{z}\}$ and the coefficient $\phi_{i,j}$ is given by the unitary transformation of $J_z$ under the SU(4) rotations. The spectral decomposition of $\hat{G}$ is given by $\hat{G}=\sum_{i=1}^{4}\lambda_{i}|\lambda_{i}\rangle\langle\lambda_{i}|$, $\lambda_{i}\in\mathbb{R}$. When $\Delta_l \ll 1$ for all $l$, the residual motional excitation is given by $D\left(\hat{F}(\boldsymbol{\Delta})\right)$  where $\hat{F}(\boldsymbol{\Delta})=\sum_{l=1}^{n}\Delta_{l}\hat{G}_{l}$ with $\hat{G}_l=\sum_{i,j}\phi_{i,j}^{(l)}\hat{S}_{i}\otimes\hat{S}_{j}$.  We assume uniform error throughout the gate sequences of the Petz recovery map e.g. $\Delta_l = \Delta$ for all $l$. The closed form of $\mE_{RM}^{(\Delta)} $ reads 
\begin{equation}
    U\left[\sum_{i,j=1}^{4}\left(e^{-(|\sqrt{\Delta}\lambda_{i}|^{2}+|\sqrt{\Delta}\lambda_{j}|^{2})/2}e^{|\Delta|\lambda_{i}\lambda_{j}}|\lambda_{i}\rangle\langle\lambda_{j}|\right)\circ\tilde{\bullet}\right]U^{\dag}
\end{equation}
where $\circ$ denotes the Hadamard product and $\tilde{\bullet} = U^{\dag}\bullet U$.
$U$ is the unitary transformation from the basis of $\hat{G}$ to the basis of $z$ (see Appendix \ref{sec:compositeDisplacement} for the full derivation). 

Lastly, we examine the recovery error between the ideal Petz recovery map and the noisy Petz recovery map
\begin{equation}
    \epsilon_{(\mE,\rho)}(\Delta) =1- \mathcal{F} \left[\hat{\mathcal{R}}_{(\mE,\rho)}\circ \mE(\rho), \hat{\mathcal{R}}_{(\mE,\rho)}^{(\Delta)}\circ \mE(\rho)\right].
\end{equation}
By uniformly sampling $\rho$ over the Bloch sphere, we give the mean recovery error in terms of $\Delta$ for the different channels in Fig. \ref{fig:FidThresAv}. We also give maximal observed value of the recovery error in Fig. \ref{fig:FidThresMax}. For the two-qubit gate error of $\Delta = 10^{-4}$ and $p=0.5$, the average recovery error is bounded below $0.01$ for all states.

\section{\label{sec:Conclusion}Conclusion}
In this paper, we have developed a systematic method to implement an exact Petz recovery map for qubit decoherence up to arbitrary Kraus-rank. Our construction is based on isometric extension and requires no post-selection. For a qubit decoherence channel, our circuit of the corresponding Petz map uses at most two ancilla qubits for single-qubit Petz recovery, which is less ancila-demanding than \cite{gilyen2022quantum}. Our circuit realisation also uses minimum amount of CNOT gate, e.g. 3 for decoherence of Kraus rank 2 and 20 for decoherence channel of Kraus rank $M>2$. In single-shot recovery protocols, we find that recovering a severely decohered high-purity state requires a more tightly constrained prior distribution of the reference state. Lastly, we situate the Petz recovery in the context of a trapped ion quantum computer, and give the error threshold of the geometric phase gate for recovery up to some finite error. For a recovery error bounded by 0.01, implementing the Petz map for a depolarizing channel requires geometric phase gate errors that are an order of magnitude lower than those required for dephasing and amplitude-damping channels. This is because the deeper circuit required by the Petz map with a higher Kraus rank is more susceptible to error accumulation.

In future work, the extension of the physical implementation of Petz recovery map to other quantum computing platforms is desired. This paves the way for realising quantum Bayesian inference on a quantum computer. In this case, multiple Petz recovery maps are required, each with updated reference state. It would be interesting to examine the exact implementation of the continuous update protocol and to examine the effectiveness of the Petz recovery map under a noisy circuit.

\begin{acknowledgments}
The authors thank Dzmitry Matsukevich, Minjeong Song, Clive Cenxin Aw for helpful discussions. This project is supported by the National Research Foundation, Singapore through the National Quantum Office, hosted in A*STAR, under its Centre for Quantum Technologies Funding Initiative (S24Q2d0009); by the Ministry of Education, Singapore, under the Tier 2 grant ``Bayesian approach to irreversibility'' (Grant No.~MOE-T2EP50123-0002), and under the Academic Research Fund Tier 1 (FY2022, A-8000988-00-00).
\end{acknowledgments}

\bibliography{ref}

\pagebreak
\onecolumngrid

\pagebreak
\appendix

\section{Appendix}

\subsection{\label{sec:PriorBoundary} Prior distribution for Petz recovery error of 0.01 }
\begin{figure}[H]
    \centering
    \includegraphics[width=0.4\linewidth]{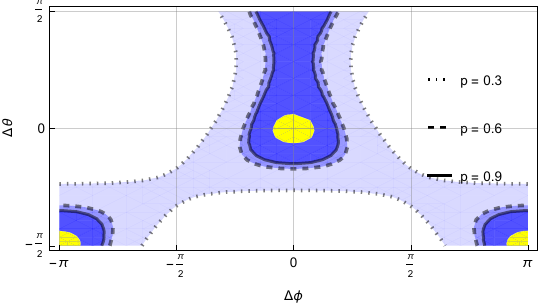} \includegraphics[width=0.4\linewidth]{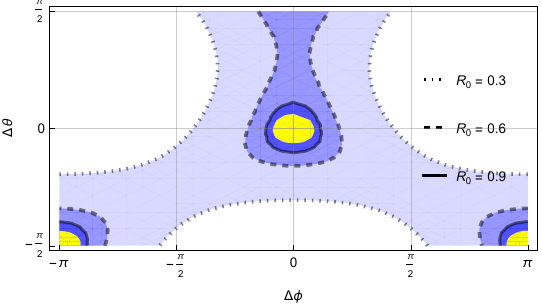}
    \includegraphics[width=0.4\linewidth]{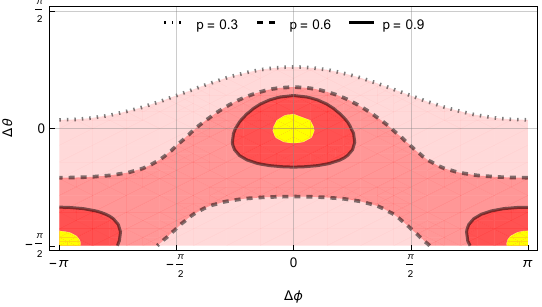} \includegraphics[width=0.4\linewidth]{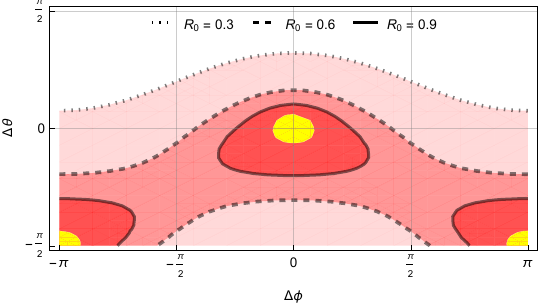}
    \includegraphics[width=0.4\linewidth]{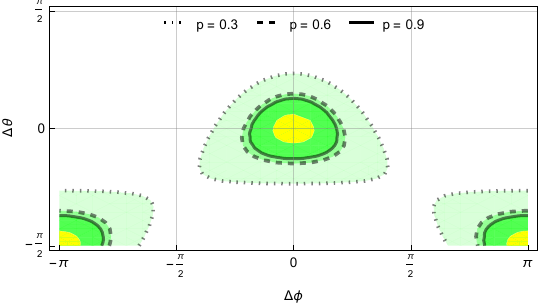} \includegraphics[width=0.4\linewidth]{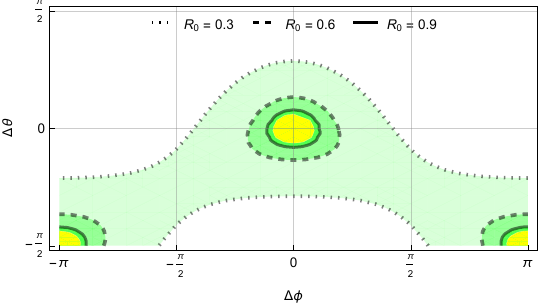}
    \caption{Parametric plot of $ \delta \mathcal{F} =0.01$ in the space of $(\Delta \phi,\Delta \theta)$. The left (right) column corresponds to $R_0 = 0.5$ ,  $p = \{0.3,0.6,0.9 \}$  ($p = 0.5$ , $R_0 = \{0.3,0.6,0.9\}$). The figure from top to down are plots for dephasing (blue), amplitude damping (red) and depolarising (green) respectively.}
    \label{fig:FidVsP0S}
\end{figure}

Here, we give the region of the prior distribution of the reference state in the 2D space of $(\Delta\theta, \Delta\phi)$ where the recovery error is equal or lower than 0.01. Using the input state with $(\theta_0 = \pi/4, \phi_0 = \pi/9)$, we give the contour line of the $\delta\mathcal{F}=0.01$  in two cases (i) $R_0 = 0.5$ ,  $p = \{0.3,0.6,0.9 \}$  (ii) $p = 0.5$ , $R_0 = \{0.3,0.6,0.9\}$  (see Fig. \ref{fig:FidVsP0S})

\subsection{\label{sec:CircuitDepo} Circuit diagram of the Petz map of Kraus rank $M=2, M=4$ }
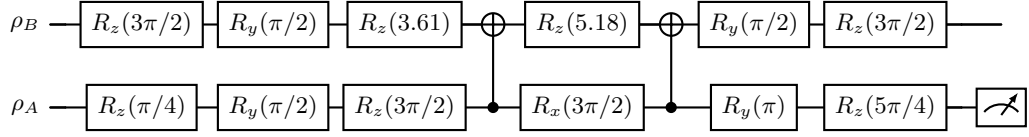
\begin{figure}[H]
\centering
\begin{quantikz}[column sep=0.2cm]
    \lstick{$\rho_B$}  & \qw  & \gate{R_z(\textnormal{$3 \pi/2$})} &\gate{R_y(\textnormal{$\pi/2$})} & \gate{R_z(\textnormal{$3.61$})} & \targ  & \qw &\gate{R_z(\textnormal{$5.18$})} & \targ & \qw &\gate{R_y(\textnormal{$\pi/2$})} &\gate{R_z(\textnormal{$3 \pi/2$})} & \qw & \qw  \\
    \lstick{$\rho_A$}  & \qw  & \gate{R_z(\textnormal{$\pi/4$})} & \gate{R_y(\textnormal{$\pi/2$})} & \gate{R_z(\textnormal{$3 \pi/2$})} & \ctrl{-1} & \gate{R_x(\textnormal{$3 \pi/2$})} & \ctrl{-1} & \gate{R_y(\textnormal{$\pi$})} &\gate{R_z(\textnormal{$5\pi/4$})} & \qw &\meter{} \\
\end{quantikz}
\caption{Gate decomposition of the Petz recovery map for dephasing channel. The rotation angles of the single-qubit gates are in radians. }
\label{Fig:CircuitDephase} 
\end{figure}

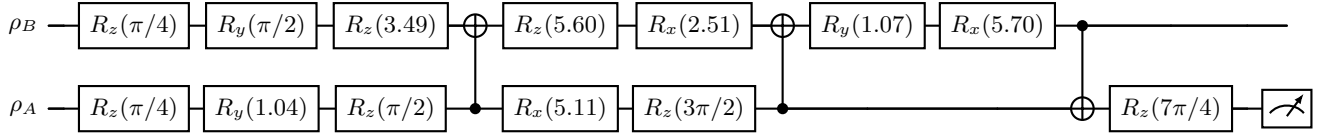
\begin{figure}[H]
\centering
\begin{quantikz}[column sep=0.2cm]
    \lstick{$\rho_B$}  & \qw  & \gate{R_z(\textnormal{$\pi/4$})} &\gate{R_y(\textnormal{$\pi/2$})} & \gate{R_z(\textnormal{$3.49$})} & \targ & \qw &\gate{R_z(\textnormal{$5.60$})} & \gate{R_x(\textnormal{$2.51$})} & \targ & \qw &\gate{R_y(\textnormal{$1.07$})} &\gate{R_x(\textnormal{$5.70$})} & \ctrl{1} & \qw  & \qw  & \qw \\
    \lstick{$\rho_A$}  & \qw  & \gate{R_z(\textnormal{$\pi/4$})} &\gate{R_y(\textnormal{$1.04$})} & \gate{R_z(\textnormal{$\pi/2$})} &\ctrl{-1} &\gate{R_x(\textnormal{$5.11$})} &\gate{R_z(\textnormal{$3 \pi/2$})} & \ctrl{-1} & \qw  & \qw  & \targ & \qw &\gate{R_z(\textnormal{$7\pi/4$})} & \qw  &\meter{} \\
\end{quantikz}
\caption{Gate decomposition of the Petz recovery map for amplitude damping channel. The rotation angles of the single-qubit gates are in radians. }
\label{Fig:CircuitAmpDamp} 
\end{figure}

After the isometric extension, we plug $U_{(\mE,\gamma)}$ into a quantum compiler \cite{iten2016quantum} for circuit decomposition. Here, we give the circuit realisation of Petz recovery for dephasing ( Fig. \ref{Fig:CircuitDephase}) and amplitude damping ( Fig. \ref{Fig:CircuitAmpDamp}), and depolarising ( Fig. \ref{Fig:CircuitDepolarising}). The example uses a reference state with $R=0.5$, $\theta = \pi/2$, $\phi = \pi/4$.

\begin{sidewaysfigure}
\centering
\begin{quantikz}[column sep=0.1cm]
     \lstick{$\rho_B$}     & \qw  & \qw  & \qw  & \qw  & \qw  & \qw  & \qw  & \qw  & \qw  & \qw  \
    &\targ &\qw & \gate{R_z} &\targ &\qw & \gate{R_z} &\targ &\qw & \gate{R_z} &\targ &\qw 
    & \gate{R_z} & \gate{R_y} & \qw  & \qw  & \qw  & \qw  & \qw  & \
    \qw  &\targ &\qw  &\qw  & \push{\cdots}\\
    \lstick{$\rho_B$} & \qw  & \gate{R_z} & \gate{R_y} & \gate{R_z} &\targ &\qw & \gate{R_z} & 
    \gate{R_x} &\targ &\qw & \gate{R_y} & \gate{R_x} & \qw  & \qw  & \ctrl{-1} 
    & \qw  & \qw  & \qw  & \ctrl{-1} & \gate{R_y} & \gate{R_z} &\targ &\qw & 
    \gate{R_z} & \gate{R_x} &\targ &\qw & \gate{R_y} & \gate{R_x} & \ctrl{-1} &\qw & \push{\cdots} \\
    \lstick{$\rho_A$}   & \qw  & \gate{R_z} & \gate{R_y} & \gate{R_z} & \ctrl{-1} & \gate{R_x} 
    & \gate{R_z} & \ctrl{-1} & \gate{R_y} & \gate{R_z} & \ctrl{-2} & \qw  
    & \qw  & \qw  & \ctrl{-2} & \gate{R_y} & \gate{R_z} & \qw  & \qw  & 
    \ctrl{-1} & \gate{R_x} & \gate{R_z} & \ctrl{-1} & \gate{R_y} & 
    \gate{R_z} & \qw &\qw & \push{\cdots}\\
\end{quantikz}

\begin{quantikz}[column sep=0.1cm]
    & \push{\cdots} & \gate{R_y} &\targ &\qw & \gate{R_y} &\targ & \qw  & \gate{R_y} & 
    \qw  & \qw  & \qw  & \qw  & \qw  & \qw  & \qw  &\targ &\qw 
    & \gate{R_z} &\targ &\qw & \gate{R_z} &\targ &\qw & \gate{R_z} &\targ &\qw & \gate{R_z} & \qw  & 
    \qw  & \qw  & \qw  & \qw  & \qw  & \qw  & \qw  & \qw  & \qw  \\
    & \push{\cdots} & \qw  & \qw  & \qw  & \ctrl{-1} & \gate{R_y} & \gate{R_z} &\targ &\qw & 
    \gate{R_z} & \gate{R_x} &\targ &\qw & \gate{R_y} & \gate{R_x} & \qw  & 
    \qw  
    & \ctrl{-1} & \qw  & \qw   & \qw  & \ctrl{-1} & \gate{R_y} & 
    \gate{R_z} &\targ &\qw & \gate{R_z} & \gate{R_x} &\targ &\qw & \gate{R_y} & 
    \gate{R_x} & \ctrl{1} & \qw  & \qw  \\
    & \push{\cdots} & \qw  & \ctrl{-2} & \gate{R_y} & \gate{R_z} & \qw  
    & \qw  & \ctrl{-1} & \gate{R_x} & \gate{R_z} & \ctrl{-1} & \gate{R_y} 
    & \gate{R_z} & \ctrl{-2} & \qw  & \qw  & \qw  & \ctrl{-2} 
    & \gate{R_y} 
    & \gate{R_z} & \qw  & \qw  & \ctrl{-1} & \gate{R_x} & \gate{R_z} & 
    \ctrl{-1} & \gate{R_y} & \gate{R_z} &\targ &\qw & \gate{R_z} & \qw  &\meter{} \\
\end{quantikz}

\caption{Gate decomposition of Petz recovery map for depolarising channel. Note that the rotation angle for each single qubit rotation are not the same here.  }
\label{Fig:CircuitDepolarising} 
\end{sidewaysfigure}


\section{Model of Noisy Petz map}

\subsection{\label{sec:ResidualDephase}Decoherence due to residual spin-motion coupling}

The elementary two-qubit  gate for trapped ion is based on the spin-dependent
motional excitation. The interaction Hamiltonian read
\[
\tilde{H}_{int}(t)=\left[A^{*}(t)\hat{a}-A(t)\hat{a}^{\dagger}\right]\hat{J}_{z}
\]

where $\hat{J}_{z}=\hat{\sigma}_{z}^{(1)}+\hat{\sigma}_{z}^{(2)}$, $A(t)=ge^{i\delta t}$
and $g$ is the coupling strength $\delta=\omega-\omega_{0}$ is the
detuning between the laser's frequency and the motional frequency.
Here we have assume that the two ions are applied with the laser of same
$g$ and $\delta$. The time-ordered evolution gives 
\begin{align*}
\mathcal{T}\exp\left(-i\int_{0}^{t}\tilde{H}(t')dt'\right) & =\exp\left(-i\left[\hat{a}\alpha^{*}(t)-\hat{a}^{\dagger}\alpha(t)\right]J_{z}\right)\exp\left(-\frac{i}{2}\Phi(t)\hat{\sigma}a_{z}^{(1)}\hat{\sigma}_{z}^{(2)}\right)\\
 & =D(\alpha(t)\hat{J}_{z})\exp\left(-\frac{i}{2}\Phi(t)\hat{\sigma}_{z}^{(1)}\hat{\sigma}_{z}^{(2)}\right)
\end{align*}

where $D(\beta) = e^{\beta a^\dag  - \beta^* a}$ and $U_{ZZ}(\theta )= \exp\left(-\frac{i}{2} \theta \hat{\sigma}_{z}^{(1)}\hat{\sigma}_{z}^{(2)}\right)$. Here,

\[
\alpha(t)=-i\frac{g}{\delta}(e^{i\delta t}-1)\sim O\left(\frac{g}{\delta}\right)
\]
\begin{align*}
\Phi(t) & =\int_{0}^{t}dt'\Im\left[A(t')\alpha^{*}(t')\right]=\frac{g^{2}}{\delta}\left(t-\frac{\sin\delta t}{\delta}\right)\sim O\left(\left[\frac{g}{\delta}\right]^{2}\right)
\end{align*}

In an ideal condition, $t=T=2\pi/\delta$, we have $\alpha(T)=0$
and $\Phi(T)=2\pi\frac{g^{2}}{\delta^{2}}$. This forms the ideal
geometric phase gate gate

\[
U_{ZZ}(\Phi(T))=\mathcal{T}\exp\left(-i\int_{0}^{T}\tilde{H}(t')dt'\right)=\exp\left(-\frac{i}{2}\Phi(T)\hat{\sigma}_{z}^{(1)}\hat{\sigma}_{z}^{(2)}\right)
\]

geometric phase gate serves as the primitive gate for CNOT up to several single-qubit rotations.  Whenever there is $t$ such that $\alpha(t)\neq0$,
this introduces the residual entanglement between the qubits and the
motion, leading to errors. Without losing generality, we consider a linear perturbation on $g$ such that $g\rightarrow g+\epsilon_g$. Since $\alpha(t)\sim O\left(\frac{g}{\delta}\right)$ and $\Phi(t)\sim O\left(\left[\frac{g}{\delta}\right]^{2}\right),$
The overall operation in terms of two-qubit gate error can be expressed in terms of $\Delta$ where $\Delta = \left[\frac{\epsilon_g}{\delta}\right]^{2} $

\[
\mathcal{T}\exp\left(-i\int_{0}^{t}\tilde{H}(t')dt'\right)=D\left(\sqrt{\Delta}\hat{J}_{z}\right)U_{ZZ}(\Delta)U_{ZZ}(\Phi(T))
\]

where the first term is the \emph{coherent error}, and the second
term is the \emph{systematic error}, and the third term is the ideal
geometric phase gate gate. For a single geometric phase gate, the overall channel can be written
as
\[
\mE_{GPG}^{(\Delta)}(\bullet)=\mE_{ph}^{(\Delta)}\circ\mathcal{U}_{sys}^{(\Delta)}\circ\mathcal{U}_{ZZ}(\bullet)
\]

$\bullet\in\mathcal{H}_{A}\otimes\mathcal{H}_{B}$ denotes the system qubit and the ancilla qubit,
and 

\begin{align*}
\mathcal{U}_{ZZ}(\bullet) & =U_{ZZ}^{\dag}(\Phi(T))\bullet U_{ZZ}(\Phi(T))\\
\mathcal{U}_{sys}^{(\Delta)}(\bullet) & =U_{ZZ}^{\dagger}(\Delta)\bullet U_{ZZ}(\Delta)\\
\mE_{ph}^{(\Delta)}(\bullet) & =\text{Tr}_{ph}\left[D\left(\sqrt{\Delta}\hat{J}_{z}\right)\left(\bullet\otimes|0\rangle\langle0|_{ph}\right)D\left( - \sqrt{\Delta}\hat{J}_{z}\right)\right]
\end{align*}

Here $ph$ denotes phonon and $|0\rangle\langle0|_{ph}$ is the motional ground state. In Kraus representation,
the residual spin-dependent motional excitation on both qubits in the specific
direction $\hat{J}_{k}=\{\hat{J}_{z},\hat{J}_{x},\hat{J}_{y}\}$  can be derived as follows

\begin{align*}
\mathcal{E}_{ph}^{(\Delta)}(\bullet) & =Tr_{B}\left[D(\sqrt{\Delta}J_{z})(\bullet\otimes|0\rangle\langle0|)D(-\sqrt{\Delta}J_{z})\right]\\
 & =\sum_{m=0}^{\infty}(I\otimes\langle m|)D(\sqrt{\Delta}J_{z})(I\otimes|0\rangle)\bullet(h.c)\\
 & =\sum_{m=0}^{\infty}\left(\begin{array}{cccc}
\langle m|D(\sqrt{\Delta})|0\rangle & 0 & 0 & 0\\
0 & 1 & 0 & 0\\
0 & 0 & 1 & 0\\
0 & 0 & 0 & \langle m|D(\sqrt{\Delta})|0\rangle
\end{array}\right)\bullet(h.c)\\
 & =e^{-|\sqrt{\Delta}|^{2}}\sum_{m=1}^{\infty}\frac{|\sqrt{\Delta}|^{2m}}{m!}\left(\begin{array}{cccc}
1 & 0 & 0 & 0\\
0 & 1 & 0 & 0\\
0 & 0 & 1 & 0\\
0 & 0 & 0 & (-1)^{m}
\end{array}\right)\bullet\left(\begin{array}{cccc}
1 & 0 & 0 & 0\\
0 & 1 & 0 & 0\\
0 & 0 & 1 & 0\\
0 & 0 & 0 & (-1)^{m}
\end{array}\right)\\
 & =e^{-|\sqrt{\Delta}|^{2}}\left[\sum_{m\in even}^{\infty}\frac{\Delta{}^{m}}{m!}\bullet+\sum_{m\in odd}^{\infty}\frac{\Delta{}^{m}}{m!}CZ\bullet CZ\right]\\
 & =e^{-|\sqrt{\Delta}|^{2}}\left[\cosh(\Delta)\bullet+\sinh(\Delta)CZ\bullet CZ\right]
\end{align*}
\\
where we have use $\langle m|\sqrt{\Delta}\rangle=e^{-|\sqrt{\Delta}|^{2}/2}\frac{(\sqrt{\Delta})^{m}}{\sqrt{m!}}$
and let $CZ=\text{diag \ensuremath{(1,1,1,-1)}}$ . In Kraus representation, 
\[
\mathcal{E}_{ph}^{(\Delta)}(\bullet)=E_{1}\bullet E_{1}^{\dag}+E_{2}\bullet E_{2}^{\dag}
\]

where

\[
E_{1}=e^{-|\sqrt{\Delta}|^{2}/2}\sqrt{\cosh(\Delta)}I
\]

\[
E_{2}=e^{-|\sqrt{\Delta}|^{2}/2}\sqrt{\sinh(\Delta)}CZ
\]

\subsection{\label{sec:NoisyPetz}Generic error from a quantum circuit due to residual spin-motion coupling}

Next, we extend the tools to calculate the errors of a quantum circuits. An ideal Petz recovery map is implemented through the following $U_{(\mE,\gamma)}$ which can be decomposed into series of single-qubit rotations and CNOT gates. Let's denote the CNOT gate with control on qubit n and target on qubit m as $CNOT(n,m)$. The decomposition of CNOT$(n,m)$ gate into primitives gates follows where $(H\delta_{1,n} \otimes H\delta_{2,m})C(Z)(H\delta_{1,n} \otimes H\delta_{2,m})$ where $\delta_{i,j}$ is a Kronecker delta, $H = (\sigma_x + \sigma_z)/\sqrt{2}$ and $C(Z) = e^{-i\pi/4} e^{i\sigma_z^{(1)} \pi/4} e^{i\sigma_z^{(2)} \pi/4} e^{-i \sigma_z^{(1)}\sigma_z^{(2)} \pi/4}$. The overall circuit of an ideal Petz map can be written as follows

\[
\mathcal{U}_{(\mE,\gamma)} (\bullet) = U_{(\mE,\gamma)} (\bullet)U_{(\mE,\gamma)}^{\dag} = (\mathcal{U}_n \circ ...\circ \mathcal{U}_1 )(\bullet)
\]

where $\mathcal{U}_n$ each is either single-qubit rotation or geometric phase gate gate. To quantify the overall circuit error due to residual spin-motion coupling, we replace every unitary channel by geometric phase gate gate in the circuit as follows $ \mE_{GPG}(\bullet)= \mE_{ideal}(\bullet) \rightarrow \mE_{GPG}(\bullet)=\mE_{ph}^{(\Delta)}\circ\mathcal{U}_{sys}^{(\Delta)}\circ\mE_{ideal}(\bullet)$. 

To illustrate the idea, we give an example model of the noisy quantum circuit of two qubits. Any two-qubit unitary $U\in SU(4)$ has decomposition as in Fig. \ref{Fig:Circuit}.

\[
U_{(\mE,\gamma)} (\bullet)U_{(\mE,\gamma)}^{\dag} = (\mathcal{U}^{(4)} \circ \mathcal{U}_{ZZ}  \circ \mathcal{U}^{(3)} \circ \mathcal{U}_{ZZ} \circ  \mathcal{U}^{(2)} \circ \mathcal{U}_{ZZ} \circ \mathcal{U}^{(1)} )(\bullet)
\]

where $ \mathcal{U}^{(n)} (\bullet) =( u_1 \otimes v_1 )\bullet ( u_1 \otimes v_1 )^\dag $ and $ \mathcal{U}_{ZZ}(\bullet) = U_{ZZ}(\pi/4) \bullet U^\dag_{ZZ}(\pi/4)$. By replacing $\mathcal{U}_{ZZ}$ with $\mE_{GPG}$, the noisy quantum circuit operation can be expressed as follows $\mE_{(\mE,\gamma)}$

\[
\mE_{(\mE,\gamma)}^{(\Delta)}  (\bullet) = (\mathcal{U}^{(4)} \circ \mE_{GPG}^{(\Delta)}  \circ \mathcal{U}^{(3)} \circ \mE_{GPG}^{(\Delta)} \circ  \mathcal{U}^{(2)} \circ \mE_{GPG}^{(\Delta)}  \circ \mathcal{U}^{(1)} )(\bullet)
\]

We can shift all $\mE_{ph}^{(\Delta)}$ to the left using the unitary transformation 

\[
\mE_{(\mE,\gamma)}^{(\Delta)} (\bullet) = \mE_{RM}^{(\Delta)} \circ \mathcal{U}_{(\mE,\gamma)}^{(\Delta)}(\bullet)
\]

where

\[
\mathcal{U}_{(\mE,\gamma)}^{(\Delta)}=  \mathcal{U}^{(4)} \circ\mathcal{U}_{sys}^{(\Delta)}\circ\mathcal{U}_{ZZ}  \circ \mathcal{U}^{(3)} \circ \mathcal{U}_{sys}^{(\Delta)}\circ\mathcal{U}_{ZZ} \circ  \mathcal{U}^{(2)} \circ \mathcal{U}_{ZZ} \circ  \mathcal{U}^{(1)}
\]
is the unitary rotation with systematic shift., and

\[
\mE_{RM}^{(\Delta)} = \mE_{ph,1}^{(\Delta)} \circ \mE_{ph,2}^{(\Delta)} \circ \mE_{ph,3}^{(\Delta)} 
\]

is the error channel due to residual spin motion entanglement with
\[
\mE_{ph,1}^{(\Delta)} = \mathcal{U}^{(4)} \circ \mE_{ph}^{(\Delta)} \circ \mathcal{U}^{(4)\dag}
\]

\[
\mE_{ph,2}^{(\Delta)} =(\mathcal{U}^{(4)} \circ\mathcal{U}_{sys}^{(\Delta)}\circ\mathcal{U}_{ZZ}  \circ \mathcal{U}^{(3)})  \circ  \mE_{ph}^{(\Delta)}  \circ  (\mathcal{U}^{(4)} \circ\mathcal{U}_{sys}^{(\Delta)}\circ\mathcal{U}_{ZZ}  \circ \mathcal{U}^{(3)})\mE_{ph}^{(\Delta)}) ^\dag
\]

\[
\mE_{ph,3}^{(\Delta)} = (\mathcal{U}^{(4)} \circ\mathcal{U}_{sys}^{(\Delta)}\circ\mathcal{U}_{ZZ}  \circ \mathcal{U}^{(3)} \circ \mathcal{U}_{sys}^{(\Delta)}\circ\mathcal{U}_{ZZ} \circ  \mathcal{U}^{(2)}) \circ \mE_{ph}^{(\Delta)}  \circ (\mathcal{U}^{(4)} \circ\mathcal{U}_{sys}^{(\Delta)}\circ\mathcal{U}_{ZZ}  \circ \mathcal{U}^{(3)} \circ \mathcal{U}_{sys}^{(\Delta)}\circ\mathcal{U}_{ZZ} \circ  \mathcal{U}^{(2)}) ^\dag
\]

Now we can define the Noisy Petz recovery map as following

\[
\hat{\mathcal{R}}_{(\mE,\gamma)}^{(\Delta)} = \text{Tr}_{B}\left[ \mE_{(\mE,\gamma)}^{(\Delta)} (\rho_{B}\otimes \bullet)\right]
\]

where $\bullet\in \mathcal{H}_A\otimes \mathcal{H}_B$. The similar modeling of noisy Petz recovery map also applicable to circuit with arbitrary number of CNOT gate

\subsection{\label{sec:compositeDisplacement}Derivation of composite spin-dependent excitation} 
One may notice that the residual spin-dependent
displacement in $\mE_{RM}^{(\Delta)}$ may not be uni-directional after unitary transformation e.g. they might not be described by $\{\hat{J}_{z},\hat{J}_{x},\hat{J}_{y}\}$. WLOG, we
consider  $D(\alpha_{k}\hat{G}_{k})$, where $\hat{G}=\sum_{i,j}\phi_{i,j}\hat{S}_{i}\otimes\hat{S}_{j}$
is a Hermitian operator that serves as a generator for two qubit operations. Here $\hat{S}_{i}=\{I,\sigma_{x},\sigma_{y,}\sigma_{z}\}$. We can
also write $\hat{G}=\sum_{i=1}^{4}\lambda_{i}|\lambda_{i}\rangle\langle\lambda_{i}|$,
$\lambda_{i}\in\mathbb{R}$. The bases $S_{i}\otimes S_{j}$ are not
orthogonal, but it is closed under SU(2) rotations. Now the spin-dependent
motional excitations read

\[
\mE_{RM}^{(\Delta)} (\bullet)= \text{Tr}_{B}\left[\Pi_{l=1}^{n}D(\sqrt{\Delta_l}\hat{G}_{l})(\bullet\otimes|0\rangle\langle0|)D(-\sqrt{\Delta_l}\hat{G}_{l})\right]
\]

where $n$ is the number of Ising gates. For two qubit unitary gates, $n =3$. The composite spin-dependent displacement is

\[
\Pi_{l}^{n}D(\sqrt{\Delta_l}\hat{G}_{l})=D\left(\sum_{l=1}^{n}\sqrt{\Delta_l}\hat{G}_{l}\right)W
\]

where $W$ a factor due to the non-commutativity between the displacement operator. In the small error
regime $\Delta_l  \ll 1,$ we use the following BCH approximation 

\begin{align*}
e^{\Delta [X,Y]} & =e^{\hat{X}\Delta }e^{\hat{Y} \Delta }e^{-\frac{\Delta ^{2}}{2}[\hat{X},\hat{Y}]}O(e^{\Delta ^{3}/6})...\\
 & \approx e^{\hat{X} \Delta }e^{ \hat{Y} \Delta}
\end{align*}

Thus, in regime of $\Delta_{l}\ll1,$ we can neglect the factor $W$.
\[
 \Pi_{l}^{n}D(\sqrt{\Delta_l}\hat{G}_{l}) \approx D\left(\sum_{l=1}^{n}\sqrt{\Delta_l}\hat{G}_{l}\right)
\]

Here
\begin{align*}
D\left(\sum_{l=1}^{n}\sqrt{\Delta_l}\hat{G}_{l}\right) & =\exp\left(\hat{F}(\boldsymbol{\alpha})\hat{a}^{\dagger}-\hat{F}^{*}(\boldsymbol{\alpha})\hat{a}\right)\\
 & =D\left(\hat{F}(\boldsymbol{\alpha})\right)
\end{align*}

where $\hat{F}(\boldsymbol{\alpha})=\sum_{l=1}^{n}\sqrt{\Delta_l}\hat{G}_{l}=\sum_{i,j}\phi'_{i,j}\hat{S}_{i}\otimes\hat{S}_{j}$ and $\phi'_{i,j} = \Delta_l \phi_{i,j}$. When the error $\Delta$ is slow varying, $\Delta_l = \Delta, \forall l$. Then we have $\hat{F}(\boldsymbol{\alpha})=\Delta\sum_{l=1}^{n}\hat{G}_{l}$,
where $\sum_{l=1}^{n}\hat{G}_{l}=\sum_{i=1}^{4}\lambda_{i}|\lambda_{i}\rangle\langle\lambda_{i}|$
, $\lambda_{i}\in\mathbb{R}$.

\begin{align*}
\mE_{RM}^{(\Delta)}(\bullet) & =U\left(\sum_{i,j=1}^{4}\sum_{m=1}^{\infty}e^{-|\sqrt{\Delta}\lambda_{i}|^{2}/2}\frac{\left(\sqrt{\Delta}\lambda_{i}\right)^{m}}{\sqrt{m!}}|\lambda_{i}\rangle\langle\lambda_{i}|\tilde{\bullet}e^{-|\sqrt{\Delta}\lambda_{j}|^{2}/2}\frac{\left(\sqrt{\Delta^{*}}\lambda_{j}\right)^{m}}{\sqrt{m!}}|\lambda_{j}\rangle\langle\lambda_{j}|\right)U^{\dag}\\
 & =U\left[\sum_{i,j=1}^{4}\left(e^{-(|\sqrt{\Delta}\lambda_{i}|^{2}+|\sqrt{\Delta}\lambda_{j}|^{2})/2}\sum_{m=1}^{\infty}\frac{\left(|\Delta|\lambda_{i}\lambda_{j}\right)^{m}}{m!}|\lambda_{i}\rangle\langle\lambda_{j}|\right)\circ\tilde{\bullet}\right]U^{\dag}\\
 & =U\left[\sum_{i,j=1}^{4}\left(e^{-(|\sqrt{\Delta}\lambda_{i}|^{2}+|\sqrt{\Delta}\lambda_{j}|^{2})/2}e^{|\Delta|\lambda_{i}\lambda_{j}}|\lambda_{i}\rangle\langle\lambda_{j}|\right)\circ\tilde{\bullet}\right]U^{\dag}
\end{align*}

where $\circ$ denotes Hadamard product and $\tilde{\bullet} = U^{\dag}\bullet U\in\mathcal{H}_{A}\otimes\mathcal{H}_{B}$,
$U$ is the unitary transformation from  basis of $\hat{G}$ to $z$- basis..

\end{document}